\documentclass{JHEP3}
\usepackage{amsmath}
\usepackage{epsfig}

\newcommand{\op}[1] {\operatorname{#1}}

\newcommand{\One} {{\bf 1}} % The unit matrix

\newcommand{\tth} {\operatorname{th}}

\newcommand{\cL} {{\cal L}}

\newcommand{\cO} {{\cal O}}
\newcommand{\al} {{\alpha}}
\newcommand{\la} {{\lambda}}
\newcommand{\La} {{\Lambda}}

\title{Superstring field theory equivalence: Ramond sector}
\author{Michael Kroyter\\ \\
Center for Theoretical Physics\\
Massachusetts Institute of Technology\\
Cambridge, MA 02139, USA\\
\\ and\\ \\
School of Physics and Astronomy\\
The Raymond and Beverly Sackler Faculty of Exact Sciences\\
Tel Aviv University, Ramat Aviv, 69978, Israel\\ \\
\email{mikroyt@mit.edu}, \email{mikroyt@tau.ac.il}}

\abstract{
We prove that the finite gauge transformation of the Ramond sector of
the modified cubic superstring field theory is ill-defined due to
collisions of picture changing operators.

Despite this problem we study to what extent could a bijective classical
correspondence between this theory and the (presumably consistent)
non-polynomial theory exist.
We find that the classical equivalence between these two theories can almost
be extended to the Ramond sector: We construct mappings between the string
fields (NS and Ramond, including Chan-Paton factors and the various GSO
sectors) of the two theories that send solutions to solutions in a way that
respects the linearized gauge symmetries in both sides and keeps the action
of the solutions invariant. The perturbative spectrum around equivalent
solutions is also isomorphic.

The problem with the cubic theory implies that the correspondence
of the linearized gauge symmetries cannot be extended to a correspondence
of the finite gauge symmetries. Hence, our equivalence is only formal, since
it relates a consistent theory to an inconsistent one.
Nonetheless, we believe that the fact that the equivalence formally works
suggests that a consistent modification of the cubic theory exists. We
construct a theory that can be considered as a first step towards a
consistent RNS cubic theory.}

\keywords{String Field Theory, Superstrings and Heterotic Strings}
\preprint{MIT-CTP-4028\\TAUP-2893-09\\NSF-KITP-09-32}

\begin{document}

%=====================
\section{Introduction}
%=====================

There are two familiar versions of (RNS covariant open) superstring
field theory\footnote{See~\cite{Fuchs:2008cc} for a review of recent
developments in string field theory. More detailed background on
the construction of superstring field theories can be found in section 8
therein.}. The first to be constructed is the modified cubic theory
of~\cite{Preitschopf:1989fc,Arefeva:1989cm,Arefeva:1989cp},
in which the NS string field carries zero picture number. This is
an improved version of Witten's theory~\cite{Witten:1986qs}, in which
collisions of picture changing operators destroy the gauge symmetry and
invaliadate the evaluation of scattering amplitudes~\cite{Wendt:1987zh}.
Despite correcting the problems of collisions of picture changing operators,
some difficulties still remain. Most familiar is the criticism against
the appearance of the picture changing operator $Y_{-2}$ in the definition
of this theory, on the ground that this operator possesses a non-trivial
kernel. It is not clear to us if this actually poses a problem, since one
can claim that the problematic string fields should not be considered as
legitimate ones\footnote{See the companion paper~\cite{Kroyter:2009zj} for
a more detailed discussions on this subject.}. However, there is also a
genuine problem with the cubic formulation: Its Ramond sector is
inconsistent, since its linearized gauge transformations cannot be
exponentiate to give finite gauge transformations, due to (again)
collisions of picture changing operators. This fact was unnoticed so far.
We prove it in section~\ref{sec:BadGauge}.

The second formulation is the non-polynomial theory, constructed by
Berkovits~\cite{Berkovits:1995ab}. This theory lives in the large Hilbert
space of the fermionized RNS superghosts~\cite{Friedan:1985ge}.
The physical degrees of freedom behave like $\xi$ times the usual
RNS degrees of freedom. Hence, the NS string field carries ghost number
zero (instead of one) and picture number zero (instead of the ``natural''
minus one picture). A novel gauge symmetry is introduced in order to
reduce the degrees of freedom by half, that is, in order to obtain
the same amount of degrees of freedom one has in the small Hilbert space.

It was recently discovered that a mapping exists between these two theories,
sending solutions of the one to solutions of the other, such that gauge
orbits are sent bijectively to gauge orbits~\cite{Fuchs:2008zx}. The
cohomologies around
solutions and the actions of solutions are invariant under this mappings.
Moreover, it was also shown in~\cite{Fuchs:2008zx} that this formalism can
be extended to include also the NS$-\,$ sectors of both
formulations~\cite{Berkovits:2000hf,Arefeva:2002mb}.
It was concluded that the theories are classically equivalent.

It should be noted that (in one direction) the mapping is
performed by a mid-point insertion on top of the field of the cubic theory.
Furthermore, a regularization had to be employed for the evaluation of the
action under the mapping and while the regularization was not explicitly
constructed, it was explained what should its properties
be (symmetry arguments). This state of affairs brings about two possible
points of view for interpreting the results of~\cite{Fuchs:2008zx}.
One option is that the mid-point insertion is a singular limit of another
family of mappings, which are more complicated but regular. If this
is the case then the theories are genuinely equivalent. The other option is
that the cubic theory should be thought of as a singular gauge limit of
the non-polynomial theory. Being a singular limit does not imply that
it is wrong. It only implies that one has to use some care when working with
it. For example, it was
realised~\cite{Kiermaier:2007jg,Kiermaier:2008jy,Kiermaier:2008qu} that
Schnabl's gauge~\cite{Schnabl:2005gv} is a singular limit of a regular
family of ``linear $b$-gauges''.
Nonetheless, it was also shown that it is possible to regularize
expressions in this gauge and get a reliable final result.

There are at least three obvious matters that should be dealt with regarding
the equivalence of the two theories. We already mentioned the somewhat
singular nature of the equivalence. The second issue is the inclusion of
the Ramond sector in the equivalence, while the third is the extension of the
mapping to the quantum level.
In fact, the second is a pre-requirement of the third, since Ramond
states can be produced in loops even when calculating processes involving
only NS fields as external states.
However, as we already mentioned, the Ramond sector of the cubic theory is
inconsistent. Hence, a genuine correspondence cannot exists.
Nonetheless, we prove the existence of a formal equivalence between the
theories in section~\ref{sec:main}. Then, in section~\ref{sec:general}, we
extend the formalism to describe arbitrary brane systems.

We believe that the cubic theory with the Ramond sector
should be thought of as being a singular limit, or a bad gauge fixing, of
another, benign theory. One candidate for the consistent theory is the
non-polynomial theory. However, the fact that the formal manipulations
performed in section~\ref{sec:main} show that (up to mid-point regularization
issues) the cubic and non-polynomial theories are formally equivalent,
seems to suggest that regardless of its inconsistency in its current
formulation, not all is bad in the cubic theory.
A possibility for defining a consistent cubic theory
is discussed in section~\ref{sec:2FermiCub}, where we study a theory with
two Ramond string fields that has some attractive features.
Nonetheless, since in this theory the Ramond sector is doubled,
it should somehow be further modified in order to give the cubic theory
that we seek. We did not find a way to do that in a manner that avoids
the infamous picture changing collisions. Hence, this theory is only
a first step towards the goal of a consistent cubic theory.
We comment on that and on some other issues in the conclusions
section~\ref{sec:conc}.

In the non-polynomial theory, the incorporation of the Ramond sector was
performed in~\cite{Berkovits:1995ab,Berkovits:2001im,Michishita:2004by}.
While this theory seems to be consistent, it is not known how
to define it covariantly using a single Ramond string field.
The best one can do covariantly is to write an action
with two Ramond fields, at pictures $\pm \frac{1}{2}$ and add a constraint
relating them.
One could try to implement the constraint using a Lagrangian multiplier.
However, this will introduce a non-trivial equation of motion for the new
Lagrangian-multiplier-string-field. It might be possible to get around this
issue by making this field a part of a quartet or by adding more Lagrangian
multipliers along the lines of~\cite{Berkovits:1997wj},
but this was not done so far.

The parity of the Ramond string fields in both theories is the same as
that of the NS string fields of the same theory.
This might come as a surprise, since the
vertex operators in the Ramond sector have an opposite parity to that
of the NS+ sector. However, since the components of the Ramond string
field represent fermions in space-time, the coefficient fields
themselves have to be odd. This brings the total parity of the Ramond
string field to the desired value, i.e., they are odd in the cubic theory
and they are even in the non-polynomial one.
We refer the reader to appendix~\ref{app:picGh} for some tables of the
quantum numbers of the relevant operators and string fields.

We end this introduction by presenting some conventions and some of
the operators we use.
Throughout this paper $[A,B]$ denotes the graded commutator, i.e.,
\begin{equation}
\label{gradCom}
[A,B]\equiv AB-(-)^{AB}BA\,,
\end{equation}
where $A$ and $B$ in the exponent represent the parity of $A$ and $B$.
An important building block in our construction is the operator,
\begin{equation}
\label{Pdef}
P(z)=-c\xi\partial \xi e^{-2\phi}(z)\,.
\end{equation}
This operator is a contracting homotopy operator for $Q$ in the large
Hilbert space, i.e.,
\begin{subequations}
\begin{equation}
\label{PQ}
[Q,P(z)]=1\,.
\end{equation}
Other important relations are,
\begin{align}
\label{Qxi}
[Q,\xi(z)]&=X(z)\,,\\
\label{EtaXi}
[\eta_0,\xi(z)]&=1\,,\\
[\eta_0,P(z)]&=Y(z)\,,
\end{align}
\end{subequations}
where $X$ and $Y$ are the picture changing operator and the inverse picture
changing operator respectively. These operators obey the OPE,
\begin{equation}
\label{XY}
X Y \sim 1\,.
\end{equation}
It is also possible to define the double picture changing operators $X_2$
and $Y_{-2}$, obeying,
\begin{equation}
\label{Ym2XY}
Y_{-2}X\sim Y\,,\qquad X_2 Y \sim X\,, \qquad X_2 Y_{-2} \sim 1\,.
\end{equation}
We will also use the following (regular) OPE's,
\begin{subequations}
\label{OPE}
\begin{align}
\label{PX}
PX &\sim \xi\,,\\
\label{xiY}
\xi Y &\sim P\,,\\
\label{PP}
P P &\sim 0\,,\\
\label{XiXi}
\xi \xi &\sim 0\,,\\
\label{Pxi}
P \xi &\sim 0\,.
\end{align}
\end{subequations}
The divergent OPE's are $XX$, $YY$, $X\xi$ and $YP$.

We will use the operators appearing in~(\ref{OPE}) as mid-point
insertions on string fields. This is not a-priori excluded, since they are
all primaries of zero conformal weight~\cite{Kroyter:2009zj}.
When these operators are inserted at the mid-point
over the string fields $\Psi_{1,2}$ having no other mid-point
insertions one finds,
\begin{equation}
\label{factorize}
(\cO_1\Psi_1)\star(\cO_2\Psi_2)=(-)^{\cO_2\Psi_1}
      (\cO_1\cO_2)(\Psi_1 \star \Psi_2)=(-)^{\cO_2(\cO_1+\Psi_1)}
      (\cO_2\cO_1)(\Psi_1 \star \Psi_2)\,,
\end{equation}
that is, the star algebra factorizes to a product of a graded-Abelian
mid-point operator-insertion algebra and a regular string field star algebra.
We shall often refer to~(\ref{OPE}), when we actually mean its
part within expressions of the form of~(\ref{factorize}).
Henceforth, we shall omit the star product, as it is the only possible
product of string fields.

%==============================================================
\section{The Ramond sector of the cubic theory is inconsistent}
%==============================================================
\label{sec:BadGauge}

Witten's cubic superstring field theory~\cite{Witten:1986qs} was shown to be
inconsistent, due to singularities in its gauge
transformations~\cite{Wendt:1987zh,Arefeva:1988nn}.
These singularities emerge from collisions of picture changing operators:
The picture changing operator $X$ is inserted at the string mid-point, which
is invariant under the star product. Hence, the double pole in the $XX$ OPE
gives rise to infinities when the linearized gauge transformation is plugged
into the action.
The origin of this problem lies in the presence of the picture changing
operator $X$ in the linearized gauge transformation in Witten's theory,
\begin{equation}
\delta A=Q\La+X[A,\La]\,.
\end{equation}
This in turn is inevitable, since the NS string field $A$ and the NS gauge
string field $\La$ are both of picture number $-1$. Changing the picture of
$\La$ to zero, would force one to introduce an insertion of the inverse
picture changing operator $Y$ on top of the $Q\La$ term and collisions would
still occur, now due to the double pole in the $YY$ OPE.

In order to remedy this problem, it was suggested that the NS string field
$A$ should have a zero picture
number~\cite{Preitschopf:1989fc,Arefeva:1989cm,Arefeva:1989cp}.
The Ramond sector is then described by the picture $-\frac{1}{2}$
string field $\al$~\cite{Preitschopf:1989fc}. The modified action reads,
\begin{subequations}
\label{cubAction}
\begin{eqnarray}
S&=& S_{NS}+S_R\,,\\
\label{cubActionNS}
S_{NS}&= &-\int Y_{-2}\big(\frac{1}{2}A Q A+\frac{1}{3}A^3\big)\,,\\
\label{cubActionR}
S_R&=& -\int Y\big(\frac{1}{2}\al Q \al+A \al^2\big),
\end{eqnarray}
\end{subequations}
and its equations of motion are,
\begin{subequations}
\label{CubEOMY}
\begin{eqnarray}
Y_{-2}(Q A + A^2) + Y \al^2 &=& 0\,,\\
Y (Q \al + [A,\al]) &=& 0\,.
\end{eqnarray}
\end{subequations}
Acting on these equations with $X_2, X$ respectively brings them to the form
\begin{subequations}
\label{CubEOM}
\begin{eqnarray}
\label{eomCubA}
Q A + A^2 + X \al^2 &=& 0\,,\\
\label{eomCubAl}
Q \al + [A,\al] &=& 0\,,
\end{eqnarray}
\end{subequations}
where~(\ref{XY}) and~(\ref{Ym2XY}) were used.

The action~(\ref{cubAction}) is invariant under the following infinitesimal
gauge transformations,
\begin{subequations}
\label{cubGauge}
\begin{eqnarray}
\delta   A &=& Q\La+[A,\La]+X[\al,\chi]\,,\\
\delta \al &=& Q\chi+[\al,\La]+[A,\chi]\,,
\end{eqnarray}
\end{subequations}
where $\La$ and $\chi$ are the NS and Ramond gauge string fields
respectively.
It is clear that no collisions can emerge when only the NS sector is
considered, since no picture changing operators appear in this case.

In~\cite{Preitschopf:1989fc}, it was claimed that this gauge transformation
is regular also in the Ramond sector. To ``prove'' that, the infinitesimal
transformation~(\ref{cubGauge})
was plugged into the equations of motion~(\ref{CubEOM}). It was found
that this does not lead to collisions of picture changing operators at the
leading order in the gauge string fields.
However, if one considers also the term quadratic with
respect to the gauge string fields, collisions do occur. One may hope that
adding the next (first non-linear) order to~(\ref{cubGauge}) produces another
singular term that cancels the former. This is plausible a-priori, since
both terms are quadratic with respect to $\chi$. Nonetheless, a direct
evaluation reveals that this is not the case.
Moreover, the singularity of the transformation can be seen even without
referring to the action. Consider for simplicity the case
$A_0=0$, $\al=\al_0\neq 0$ and act with a gauge transformation in the Ramond
sector, i.e., take $\La=0$ and $\chi\neq 0$. Further, assume for simplicity
that $Q\chi=0$.
Explicit iterations of the linearized gauge transformations give,
\begin{subequations}
\label{itLinGauge}
\begin{align}
\al &\rightarrow \quad\:\al_0\hspace{0.5cm}\rightarrow
 \al_0 +X\big[[\al_0,\chi],\chi\big]\hspace{0.14cm}
  \rightarrow\hspace{0.14cm}\ldots\,,\\
A & \rightarrow X[\al_0,\chi]\rightarrow \hspace{0.84cm}
 2X[\al_0,\chi] \hspace{0.72cm}\rightarrow
 3X[\al_0,\chi]+ X^2\Big[\!\big[[\al_0,\chi],\chi\big],\chi\Big].
\end{align}
\end{subequations}
We see that the expression we get for $A$ after the third iteration is
generically divergent. Furthermore, by plugging this expression into the
action and expanding it to second order with respect to $\chi$ we also obtain
divergences, as stated.

One may think that iterating the linearized transformation, as we do here, is
too naive and that the full non-linear transformation will somehow manage to
avoid this problem. This is not the case. The full non-linear transformation
is obtained from iterating the linearized one, while rescaling the gauge
fields at the $n^{\tth}$ iteration as
\begin{equation}
\La\rightarrow \frac{\la}{n}\La\,,\qquad \chi\rightarrow \frac{\la}{n}\chi\,,
\end{equation}
where $\la$ is a fixed parameter.
This fixes the coefficient of the linearized transformation to $\la$, while
assuring the infinitesimal character of the gauge field. Hence, the only
difference between the expression above and the exact one is in the
numerical values of the coefficients, which will not change dramatically.
Nonetheless, in order to remove any doubts let us consider also the full
non-linear transformation, pretending for the moment that it exists.
The differential equation defining it is of the form,
\begin{equation}
\label{difGaugeEq}
\frac{d}{d\la} \vec A(\la)=V+\cL\vec A(\la)\,.
\end{equation}
Here, we defined,
\begin{equation}
\vec A(\la)\equiv
 \left(\begin{array}{c}A(\la)\\ \al(\la)\end{array} \right).
\end{equation}
Now, $\la$ serves as an evolution parameter for the gauge transformation
and the initial condition $\vec A(0)$ is the string field before the
gauge transformation. The fixed string field $V$ and linear
operator $\cL$ in our case are,
\begin{align}
V &=\left(\begin{array}{c}Q\La\\ Q\chi\end{array} \right),\\
\cL &=\left(
  \begin{array}{cc}
      \La_R-\La_L & X(\chi_R-\chi_L)\\
      \chi_R-\chi_L & \La_R-\La_L
  \end{array} \right),
\end{align}
where $\La_R$ and $\chi_R$ represent multiplication from the right by
$\La$ or $\chi$, while $\La_L$ and $\chi_L$ operate from the left.
Note, that left and right operations commute.
The general solution of~(\ref{difGaugeEq}) is,
\begin{equation}
\label{gaugeSol}
\vec A=e^{\la\cL}\vec A(0)+\frac{e^{\la\cL}-1}{\cL}V\,,
\end{equation}
where the operator multiplying $V$ is defined by its Taylor series and is
regular.
It is easy to verify that in the pure NS case~(\ref{gaugeSol}) reduces to
the familiar form,
\begin{equation}
A(\la)=e^{-\la\La}\big(A(0)+Q \big)e^{\la\La}\,.
\end{equation}
For the example considered in~(\ref{itLinGauge}), we see that the divergent
term $X^2\Big[\!\big[[\al_0,\chi],\chi\big],\chi\Big]$ indeed appears and its
coefficient is $\frac{\la^3}{6}$. Higher order terms have higher degree of
divergence, i.e., higher powers of $X$, multiplying other conformal fields
and the total expression has the form of an essential singularity.

A way out could have still existed. The powers of $X$ in~(\ref{gaugeSol})
are partially correlated with the power of the gauge string field $\chi$.
Hence, constraining $\chi$ to always carry a factor of $Y$ in its definition
could potentially eliminate the singularity. However, this will introduce
singularities due to collisions of $Y$ in~(\ref{gaugeSol}).
In fact, in this case singularities will emerge even earlier, since the
action also contains a factor of $Y$. An operator that could have worked is
$P$, since its OPE with $X$ give $\xi$, which has a trivial OPE with $P$
and on the other hand, no singularities can emerge from iterations of $P$
itself~(\ref{OPE}). Nonetheless, this cannot be an accepted resolution, since
$P$ and $\xi$ do not live in the small Hilbert space, to which $\vec A$
belongs. Replacing $P(i)$ by, say, $P(i)-P(0)$ that does live in the small
Hilbert space would not solve the problem, since some of the singularities
would still be left. The only option for eliminating the singularities in
such a way would be to use $P(i)-P(-i)$. This, however, does not resolve the
problem of singularities in the action, since the $YP$ OPE diverges.
Moreover, if one constrains the gauge string field to contain some given
mid-point insertion, it seems to us that the physical string field should
also be constrained in a similar way, since otherwise the linearized equation
of motion would not correspond to the world sheet theory results.
Constraining the physical string field to have some insertions as part of
its definition essentially leads to a redefinition of the theory.
The redefined theory can be Witten's theory or some other variant,
e.g.,~\cite{Lechtenfeld:1988tr}.
However, it seems that there is no simple modification of the mid-point
structure that resolves the
singularities both in the gauge transformation and in the action.

Another possible resolution would be to replace the $X$ in the definition of
the linearized gauge transformation by some sort of an operator that will
effectively induce a projection to the correct space of string
fields\footnote{This possibility was proposed to me by Scott Yost.}.
It might be the case that a variant of the regularized $X$ insertions
of~\cite{Preitschopf:1989fc} would work. These variants are non-local
operators, that were introduced in order to resolve the singularities in
the tree-level scattering amplitudes. Presumably, if they are good for
resolving one sort of a singularity of the theory they might also help
to resolve the other. However, for the sake of the linearized gauge
transformation, the mid-point character of the $X$ insertion is important.
Specifically, the relation,
\begin{equation}
X(\Psi_1\Psi_2)=(X \Psi_1)\Psi_2=\Psi_1(X\Psi_2)\,,
\end{equation}
is imperative for proving that this is actually a symmetry.
It is not clear to us how could a non-local variant of $X$ obey this
relation. Also, one would still have to verify that the resulting (finite)
gauge symmetry is singularity-free. These points would have to be addressed
in any attempt to resolve the gauge symmetry singularities along these lines.
Such a resolution, if possible at all, would be a non-trivial redefinition of
the gauge symmetry.

We conclude that, for the current formulation of the theory, the well-defined
(up to issues regarding the space of string fields) linearized gauge
transformation cannot be exponentiated to a finite gauge transformation.
Hence, the fermionic gauge invariance is lost.
One might be tempted to give up this invariance. However, this would
result in a wrong number of fermionic degrees of freedom, the theory
is then no longer open string theory and there is no reason to
believe that it would be well defined quantum mechanically.
Hence, the Ramond part of the cubic action in its current form cannot be
trusted.

%========================
\section{The equivalence}
%========================
\label{sec:main}

Regardless of the inconsistency of the cubic theory, we would like to
examine to what extent could the NS sector equivalence of~\cite{Fuchs:2008zx}
be extended to the Ramond sector. We manage to construct mappings
between solutions of both theories in~\ref{sec:mainMap} and we prove
in~\ref{sec:action} that (up to the question of the existence of a
regularization) the action of corresponding solutions is the same.
All that is still needed for establishing a classical equivalence is to prove
that gauge orbits are sent to gauge orbits bijectively. This cannot be quite
the case, since in the case of the cubic theory the gauge orbits are not well
defined. Nonetheless, we can check these statements at the linearized level,
where the cubic theory does not suffer from any obvious problems.
We find in~\ref{main:gauge} that the correspondence indeed holds at that
level. This also implies that the cohomologies around equivalent solutions
are the same.

The cubic theory was already presented in section~\ref{sec:BadGauge}.
Let us now turn to the non-polynomial theory, due to
Berkovits~\cite{Berkovits:1995ab}. In this theory picture changing
operators are avoided altogether by working in the large Hilbert
space. This enables one to work with a string field of ghost and picture
number zero in the NS sector.
The doubling of the degrees of freedom is compensated by a
novel gauge symmetry.
Again, the action can be written as a sum,
\begin{subequations}
\label{NPaction}
\begin{equation}
S=S_{NS}+S_R\,.
\end{equation}
The NS part of the action was given
in~\cite{Berkovits:1995ab},
\begin{equation}
\label{BerkoAction}
S_{NS}=\frac{1}{2}\oint\Big(
e^{-\Phi} Q e^\Phi e^{-\Phi} \eta_0 e^\Phi-\int_0^1 dt\,
   \Phi [e^{-t\Phi}\eta_0 e^{t \Phi},e^{-t\Phi} Q e^{t \Phi}]
\Big),
\end{equation}
where the integral $\oint$ represents integration in the large
Hilbert space\footnote{A novel representation of this action was given
in~\cite{Berkovits:2004xh}, $S_{NS}=-\oint\!\int_0^1 dt\, \eta A_t A_Q$,
where, $A_\nabla$ stands for $e^{-t\Phi} \nabla e^{t\Phi}$ and $\nabla$
is an arbitrary derivation of the star product. Here, $\nabla$ can represent
one of the odd canonical derivations $Q$ and $\eta_0$, as well as the
variation $\delta$ or the derivative with respect to the parameter $t$,
$\partial_t$. To get from~(\ref{BerkoAction}) to this form one has to
rewrite the former as,
$S_{NS}=\int_0^1 dt\,\oint\Big(\partial_t(A_Q A_\eta)-A_t[A_\eta,A_Q]\Big)$
and use several times the identity,
$F_{\nabla_1 \nabla_2}\equiv \nabla_1 A_{\nabla_2}
            -(-)^{\nabla_1 \nabla_2}\nabla_2 A_{\nabla_1}
            +[A_{\nabla_1},A_{\nabla_2}]=0\,.$}.

It is not easy to add the Ramond sector to the non-polynomial theory. The
equations of motion were found in~\cite{Berkovits:1995ab}, but it seems that
they cannot be derived from a covariant action.
In~\cite{Michishita:2004by}, the difficulty with the Ramond sector was
attributed to a self duality property of the string field\footnote{The
Ramond string field contains a massless chiral fermion.}. It was suggested
there, in analogy with the treatment of the type IIB
self dual RR form, to introduce
another string field and impose a self duality constraint between the two
Ramond string fields. The Ramond part of the action is then,
\begin{equation}
\label{NPactionR}
S_R=-\frac{1}{2} \oint e^{-\Phi} Q \Xi e^\Phi \eta_0 \Psi\,,
\end{equation}
where $\Xi$ and $\Psi$ are the Ramond string fields. The action should be
supplemented with the constraint,
\begin{equation}
\label{constraint}
e^{-\Phi} Q\Xi e^\Phi = \eta_0 \Psi\,,
\end{equation}
\end{subequations}
which should be imposed only after deriving the equations of
motion.

The equations of motion are now,
\begin{subequations}
\label{NPeqNoConstr}
\begin{eqnarray}
\label{NPeqPhi}
\eta_0 (e^{-\Phi} Q e^\Phi) +
 \frac{1}{2}[\eta_0 \Psi, e^{-\Phi} Q \Xi e^\Phi]&=&0\,,\\
\label{NPeqXi}
\eta_0 (e^{-\Phi} Q\Xi e^\Phi) &=& 0\,,\\
\label{NPeqPsi}
Q (e^{\Phi} \eta_0 \Psi e^{-\Phi}) &=& 0\,.
\end{eqnarray}
\end{subequations}
Taking the constraint~(\ref{constraint}) into consideration this set of
equations reduces to,
\begin{subequations}
\label{NPEOM}
\begin{eqnarray}
\eta_0 (e^{-\Phi} Q e^\Phi) + (\eta_0 \Psi)^2&=& 0\,,\\
Q (e^{\Phi} \eta_0 \Psi e^{-\Phi}) &=& 0\,.
\end{eqnarray}
\end{subequations}

%=======================
\subsection{The mapping}
%=======================
\label{sec:mainMap}

For the NS sector, the mapping is given by
\begin{equation}
\label{mapA}
\tilde A=e^{-\Phi} Q e^\Phi\,.
\end{equation}
We append this mapping with the Ramond counterpart,
\begin{equation}
\label{mapAl}
\al=i \eta_0 \Psi\,.
\end{equation}
In these variables the equations of motion~(\ref{NPEOM}) take the form,
\begin{subequations}
\begin{eqnarray}
\label{AnonSmallH}
\eta_0 \tilde A - \al^2&=& 0\,,\\
\label{alEOM}
Q \al +[\tilde A, \al] &=& 0\,.
\end{eqnarray}
These two equations should be appended with the consistency conditions
that follow from the definitions~(\ref{mapA}) and~(\ref{mapAl}),
\begin{eqnarray}
\label{Aeom}
Q\tilde A+\tilde A^2 &=& 0\,,\\
\label{alSmall}
 \eta_0 \al &=& 0\,.
\end{eqnarray}
\end{subequations}
The last equation implies that $\al$ is defined in the small Hilbert space.
However, the introduction of a non-trivial Ramond field implies that
$\tilde A$ is no longer a member of the small Hilbert space, as can be read
from~(\ref{AnonSmallH}).
To remedy this problem we define,
\begin{equation}
A=\tilde A-\xi \al^2\,.
\end{equation}
This string field does live in the small Hilbert space as a result of the
definition~(\ref{AnonSmallH}) and the commutation relation~(\ref{EtaXi}).

Now, both variables live in the small Hilbert space as a result
of~(\ref{AnonSmallH}) and~(\ref{alSmall}) and the role of the equations
of motion is played by~(\ref{alEOM}) and~(\ref{Aeom}).
Rewriting these equations in terms of $A,\al$ we get exactly the
equations of motion of the cubic theory~(\ref{CubEOM}).
To that end, one has to use the nilpotency property~(\ref{XiXi})
as well as the fact that $\xi$ is a midpoint insertion.
To summarize, our mapping is given by,
\begin{subequations}
\label{Map}
\begin{eqnarray}
\label{A}
A &=& e^{-\Phi} Q e^\Phi + \xi (\eta_0 \Psi)^2\,,\\
\label{alOfPsi}
\al &=& i \eta_0 \Psi\,.
\end{eqnarray}
\end{subequations}
We discuss why the NS sector mapping of~(\ref{A}) has to be modified in
appendix~\ref{App:Why}.

For the inverse mapping we define,
\begin{subequations}
\label{invMap}
\begin{eqnarray}
\label{PhiofAmap}
\Phi &=& P A \,,\\
\label{PsiofAl}
\Psi &=& -i \xi \al\,.
\end{eqnarray}
These definitions are enough for showing that the equations of
motion are invariant. However, since we are also interested in proving the
invariance of the action, we also add the definition,
\begin{equation}
\ \ \Xi=-iP\al\,.
\end{equation}
\end{subequations}
One can check that the definitions of $\Psi$ and $\Xi$ are consistent with
the constraint~(\ref{constraint}).

Composing the maps~(\ref{Map}) and~(\ref{invMap}) one gets,
\begin{subequations}
\begin{eqnarray}
\label{Anew}
A_{\op{new}}&=& \, e^{-\Phi} Q e^\Phi-\xi \al^2=(1-PA)Q(PA)-\xi \al^2\\
\nonumber
  &=& \, A-P(QA+A^2)-\xi \al^2=A+P X \al^2 -\xi \al^2=A\,,\\
\label{alNew}
\al_{\op{new}}&=&\eta_0 \xi \al=\al\,.
\end{eqnarray}
\end{subequations}
In~(\ref{Anew}) we used the nilpotency of $P$~(\ref{PP}),
the equation of motion~(\ref{eomCubA}), as well as the OPE~(\ref{PX}),
while in~(\ref{alNew}) we used the fact that $\al$ is defined in the small
Hilbert space.
In the opposite direction one gets,
\begin{subequations}
\label{newAPhi}
\begin{eqnarray}
\label{PhiofA}
e^\Phi_{\op{new}}&=& 1+P\big(-Q e^{-\Phi} e^\Phi-\xi \al^2\big)=
     Q P e^{-\Phi} e^\Phi=e^{-QP\Phi} e^\Phi\,,\\
\label{NewPsi}
\Psi_{\op{new}}&=& \xi \eta_0 \Psi=\Psi-\eta_0 \xi \Psi\,,
\end{eqnarray}
\end{subequations}
where in~(\ref{PhiofA}) we first used the OPE~(\ref{Pxi}) and then we
used~(\ref{PP}) and (\ref{PQ}).
In this direction the composition of the transformations gives the identity
operator only for a specific gauge choice for $\Psi$. Otherwise, it gives a
gauge equivalent string field as will be shown in~\ref{main:gauge}.

%======================
\subsection{The action}
%======================
\label{sec:action}

We want to prove that the action of solutions is the same in both
theories. For the NS part, this was shown to hold in~\cite{Fuchs:2008zx}.
In fact, what was proven there is the following: Given the
mapping~(\ref{PhiofAmap}), the values of the actions~(\ref{cubActionNS})
and~(\ref{BerkoAction}) are the same. For this proof the equations of motion
were not used. Hence, we can use it here, regardless of the fact
that the equations of motion of the NS sector are modified by Ramond terms.

To complete the proof we now show that in both theories the Ramond part of
the action of an arbitrary solution is zero.
For the cubic theory this results from the fact that both terms of the
Ramond part of the action are proportional to $\al^2$ and hence to the star
product of $\al$ with the $\al$ equation of motion~(\ref{eomCubAl}).
For the non-polynomial theory,~(\ref{NPeqXi}) implies that the Ramond part
of the action integrand is annihilated by $\eta_0$. Hence,
its large-Hilbert-space integral is zero.

While we proved that the Ramond sector poses no new problems for the
equality of the action, one should remember that for the proof of
equality in the NS sector, the existence of an adequate regularization
should be assumed~\cite{Fuchs:2008zx}. We again assume that such a
regularization exists.

%=================================
\subsection{Gauge transformations}
%=================================
\label{main:gauge}

As we already stressed, the finite gauge transformation of the cubic theory
is not well defined. Hence, strictly speaking, there is no way to match gauge
orbits between the two theories.
Nevertheless, when restricted to linearized gauge transformations, all
expressions make sense. We would therefore like to study the
linearized gauge transformations,
assuming that the action~(\ref{cubAction}) is some sort of a
singular limit of a well behaved cubic theory. Since the singularity
of the cubic theory does not expresses itself at the linearized level, we
expect that, at that level, the formal equivalence that we study works.

For the gauge transformations on the side of the non-polynomial theory,
one should distinguish between the gauge symmetries of the
action~(\ref{NPaction}) and that of the equations
of motion~(\ref{NPEOM}), which is obtained after the use of the
constraint~(\ref{constraint}).
For the action, the gauge symmetries are,
\begin{subequations}
\label{actionGauge}
\begin{eqnarray}
\delta e^\Phi &=& e^\Phi \eta_0 \La_1+Q\La_0 e^\Phi\,,\\
\delta \Psi &=& \eta_0 \La_{\frac{3}{2}}+[\Psi,\eta_0\La_1]\,,\\
\delta \Xi &=& Q \La_{-\frac{1}{2}}+[Q\La_0,\Xi]\,,
\end{eqnarray}
\end{subequations}
where the four gauge string fields are labeled by their picture numbers.
This symmetry is consistent with the constraint~(\ref{constraint}).
With this constraint imposed, the gauge transformation generated by
$\La_{-\frac{1}{2}}$ becomes trivial. On the other hand, on the constraint
surface there is an enhancement of symmetry, i.e., a new gauge string field
$\La_{\frac{1}{2}}$ generates gauge transformations on this
surface~\cite{Michishita:2004by}.
The gauge symmetry takes the form\footnote{Note that $\La_0$ here is
$-\La_Q$ of~\cite{Fuchs:2008zx}.},
\begin{subequations}
\label{EOMgaugeSym}
\begin{eqnarray}
\delta e^\Phi &=& e^\Phi (\eta_0 \La_1-[\eta_0\Psi,\La_{\frac{1}{2}}])
     +Q\La_0 e^\Phi\,,\\
\delta \Psi &=& \eta_0 \La_{\frac{3}{2}}+[\Psi,\eta_0\La_1]+
      Q\La_{\frac{1}{2}}+[e^{-\Phi} Q e^\Phi,\La_{\frac{1}{2}}]\,.
\end{eqnarray}
\end{subequations}

Suppose now that we map a solution of the cubic theory to the non-polynomial
one and that this cubic solution is modified by a gauge transformation.
This induces the following transformation on the side of the non-polynomial
theory, 
\begin{subequations}
\label{gaugeCubNP}
\begin{eqnarray}
\delta \Phi &=& P\delta A=P(Q\La+[A,\La]+X[\al,\chi])\,,\\
\delta \Psi &=& -i\xi \delta \al = -i \xi (Q\chi+[\al,\La]+[A,\chi])\,.
\end{eqnarray}
\end{subequations}
The map~(\ref{PhiofAmap}) together with the nilpotency of P~(\ref{PP})
implies that
\begin{subequations}
\begin{eqnarray}
e^\Phi &=& 1+\Phi=1+PA\,,\\
\delta e^\Phi &=& \delta\Phi\,.
\end{eqnarray}
\end{subequations}
Now, define
\begin{equation}
\La_0=-P\La\,,\qquad \La_{\frac{1}{2}}=i\xi \chi\,,\qquad \La_1=\xi
  \La\,,\qquad \La_{\frac{3}{2}}=-i \tilde \xi X \chi\,.
\end{equation}
Here, $\tilde \xi$ in the definition of $\La_{\frac{3}{2}}$
represents a $\xi$ insertion at any arbitrary point other than the midpoint,
in order to avoid singularities from the OPE of $X$ and $\xi$.
Alternatively, we can take the normal ordered product $:\!\!\xi X\!\!:$.
Since $\La_{\frac{3}{2}}$ appears only in the combination
$\eta_0 \La_{\frac{3}{2}}$, the point at which $\tilde \xi$ is
inserted is of no consequence.
With these gauge string fields the transformation~(\ref{EOMgaugeSym}) takes
the form
\begin{subequations}
\begin{eqnarray}
&\phantom{.}& \begin{aligned}
\delta \Phi &= (1+P A) (\eta_0 \xi \La-[\eta_0 \xi \al,\xi \chi])
     -Q P\La (1+P A)\\
     &= (1+P A) (\La-[\al,\xi \chi])
     +(PQ \La-\La) (1+P A)\\
     &= \La+P A\La+\xi[\al,\chi]+PQ \La-\La-P\La A\,,
\end{aligned}
\\
&\phantom{.}& \begin{aligned}
\delta \Psi &= -i \eta_0 \tilde \xi X\chi-i[\xi\al,\eta_0 \xi \La]+
      i Q \xi \chi+i[(1-P A) Q (1+P A),\xi \chi]\\
      &= -i X\chi-i\xi[\al,\La]+i X\chi-
      i \xi Q \chi-i\xi[A, \chi] \,.
\end{aligned}
\end{eqnarray}
\end{subequations}
These transformations coincide with those of~(\ref{gaugeCubNP}).
Hence, a gauge transformation of the cubic theory induces a gauge
transformation of the non-polynomial theory.

Let there now be two gauge equivalent solutions of the non-polynomial theory.
There are four gauge string fields relating these two solutions,
$\La_0,\ \La_{\frac{1}{2}},\La_1,\ \La_{\frac{3}{2}}$.
It is easy to see that $\La_0$ and $\La_{\frac{3}{2}}$ do not induce any
variation of $A$ and $\al$. As for $\La_{\frac{1}{2}}$ and $\La_1$, one
can see that they induce a gauge transformation, where the gauge string
fields in the side of the cubic theory are given by,
\begin{equation}
\chi=-i\eta_0 \La_{\frac{1}{2}}\,,\qquad
\La=\eta_0\La_1+[\al,\chi]\,.
\end{equation}
The proof is similar to the above, even if somewhat longer, and makes
use of the equation of motion~(\ref{eomCubAl}), the OPE~(\ref{XiXi}),
the relation~(\ref{Qxi}) and the graded Jacobi
identity\footnote{Recall that $[\cdot,\cdot]$ is the graded commutator in our
notations~(\ref{gradCom}).} for the fields $\al,\chi$ and $A$,
\begin{equation}
[A,[\al,\chi]]-[\al,[\chi,A]]+[\chi,[A,\al]]=0\,.
\end{equation}

Finally, we have to show that mapping a solution of the non-polynomial theory
to the cubic theory and then back to the non-polynomial one results in a
solution, which is gauge equivalent to the original one. The opposite
assertion is trivial, since as we showed, composing the mappings in the
opposite order results in the identity mapping. Now, we have to use the
finite form of the gauge transformation, since the original solution and the
one obtained after the mappings are by no means infinitesimally close.
There is no problem in working with the finite gauge transformation, since
we now consider the side of the non-polynomial theory, where the finite
gauge transformation is well defined.
The expressions for the fields after the mappings are given
by~(\ref{newAPhi}). It is clear that we do not have to exponentiate the full
linearized gauge symmetry~(\ref{EOMgaugeSym}).
All that is needed is to consider the
following non-zero gauge fields,
\begin{equation}
\La_0=-P\Psi\,,\qquad \La_{\frac{3}{2}}=-\xi\Psi\,.
\end{equation}
The finite gauge transformation generated by these fields gives
exactly~(\ref{newAPhi}).

We can now conclude that (on-shell and up to the problems in the Ramond
sector of the cubic theory) gauge orbits in one theory correspond to
gauge orbits in the other theory. Also, the fact that the equations of
motion and gauge symmetries are mapped to each other in both directions
implies that the same holds also for their linearized versions. Hence,
both theories have the same cohomologies around solutions.

%===========================================================
\section{The equivalence for general D-brane configurations}
%===========================================================
\label{sec:general}

The general D-brane system introduces the need for Chan-Paton factors, as
well as the choice of sectors (NS$\pm$ / R$\pm$) that enter into each
entry of the Chan-Paton matrix\footnote{Here, we consider explicitly
the ten dimensional, flat, Poincar\'e-invariant cases. Generalization to the
case with lower dimensional D-branes should be simple.}. 
The study of the possible Chan-Paton factors and NS/R sectors is
nothing but the classification of possible open string
theories\footnote{See~\cite{Bianchi:1990yu} for a thorough discussion on
open string classification.}.
For this classification we need to impose the requirements of mutual
locality and closure of all the OPE's, as well as the consistency of
the interaction with closed strings.
The requirement of local OPE's of all fields involved,
implies that the NS$-$ sector cannot exist in the same Chan-Paton entry
with Ramond fields and that the $R\pm$ fields are also mutually exclusive.
Another requirement is the closure of the OPE, which implies that the NS+
sector is always present. Hence (at any given Chan-Paton entry) one may have
either only the NS+ sector,
or the NS+ sector with NS$-$ or with one of the R sectors.
The combination of NS+ with either of R$\pm$ can be realised on the D-brane
and on the \= D-brane. The NS$\pm$ case is realised on the non-BPS D-brane.
The pure NS+ case can also be realised~\cite{Thorn:2008ay}.

The introduction of Chan-Paton factors
into string field theory is easy. One simply tensors each string field with
the appropriate Chan-Paton matrix and adds to the definition of the integral
also a normalized trace over the Chan-Paton space~\cite{Berkovits:2000hf}.
The operators of the theories, namely $Q$, $\eta_0$ and $Y_{-2}$ are not
affected and can be thought of as being multiplied by the identity matrix in
the Chan-Paton space.

Adding the NS$-$ sector to the NS+ one, when working with a non-BPS
D-brane (or at an off-diagonal entry of the Chan-Paton matrix that represents
strings stretching between a D-brane and a \= D-brane), can be achieved by
tensoring the previous structure also with ``internal Chan-Paton''
(two by two) matrices and adding to the definition of the integral also a
normalized trace over this sector\footnote{The internal Chan-Paton matrices
compensate for the opposite Grassmann parity of the two sectors and restrict
the interaction terms only to those that respect the GSO parity.}.
This structure was first introduced for the non-polynomial theory
in~\cite{Berkovits:2000hf}, where it was shown that the NS+ sector should
be tensored with the two by two identity matrix and the NS$-$ sector
should be tensored with the Pauli matrix $\sigma_1$.
The gauge string fields for these sectors are tensored with $\sigma_3$
and $i\sigma_2$ respectively and the operators $Q$ and $\eta_0$ are also
tensored with $\sigma_3$.
For the cubic theory this structure was introduced in~\cite{Arefeva:2002mb},
where it was shown that the roles of the string fields and the gauge fields
are revered, e.g., the NS$-$ string field $A_-$ gets the $i\sigma_2$ factor
and the NS$-$ gauge field gets the $\sigma_1$. The kinetic operator $Q$
retains the $\sigma_3$ factor that should also be granted to $Y_{-2}$.

This similarity in the structures of describing the NS$\pm$ sectors in
both theories, makes the generalization of the mapping to the NS$-$ sector
straightforward. Indeed, the generalization of the mapping for the NS$-$
sector and for the case of Chan-Paton factors was already given
in~\cite{Fuchs:2008zx}. All that is needed is to tensor $\xi$ and
$P$ with $\sigma_3$ (and with the identity matrix in the genuine
Chan-Paton space) and the mappings work.

Now, we want to consider the case of adding also R$\pm$ sectors in
various entries of the Chan-Paton matrix.
As mentioned in the introduction, the Ramond string fields are odd, just
like the NS+ string field. Furthermore, we never deal with both Ramond
sectors or with a Ramond sector together with an NS$-$ sector. Hence, there
is no need to append any of the Ramond sectors with internal Chan-Paton
factors. Nevertheless, in the case where some other entries of the
Chan-Paton matrix contain an NS$-$
sector, it is possible, just for the sake of a uniform description of the
whole Chan-Paton matrix, to append internal Chan-Paton factors also to the
Ramond sector string fields. In such a case one should define an internal
Chan-Paton factor for the Ramond string field, $\sigma_R$.
The $\sigma_3$ factors on $A$, $Q$ and $Y_{-2}$ imply that $\sigma_R$ should
square to the identity matrix and commute with $\sigma_3$. Hence, one should
either choose $\sigma_R=\sigma_3$ or $\sigma_R=\One$.

The same can be done for the non-polynomial theory. There, the string fields
$\Xi$ and $\Psi$ would have to be appended with $\sigma_R$ that obeys the
same conditions as in the case of the cubic theory. Now, commutativity
with $\sigma_3$ should be implied due to its presence in $Q$ and $\eta_0$.

For our mapping to work in these cases as well, one should make the
opposite choice for $\sigma_R$ in the two theories, since $\sigma_3$
appears with an odd power everywhere in our mappings~(\ref{Map})
and~(\ref{invMap}). The most natural definition would be to define
$\sigma_R=\One$ in the non-polynomial theory and $\sigma_R=\sigma_3$
in the cubic theory. Then, all the string fields (other than the NS$-$
ones) at any of the two theories, are appended with the same factor.
The same uniformity holds also for the gauge string fields.
All the good properties of our mappings are maintained.

%============================================================
\section{A cubic action with two fields in the Ramond sector}
%============================================================
\label{sec:2FermiCub}

Establishing the correspondence between the cubic theory~(\ref{cubAction})
and the non-polynomial theory~(\ref{NPaction}) including the
constraint~(\ref{constraint}), one may ask whether there is some
sort of an extension of the mapping for the unconstrained non-polynomial
theory as well.
For such a mapping to exist, we have to consider a modification
of the cubic theory with two Ramond fields. In fact, this is a natural
avenue from the perspective of the cubic theory as well, as we
explain below.

Furthermore, the fact, discussed above, that the Ramond sector of the cubic
theory is ill-defined, calls for a refined, mid-point-insertion-free
formulation. Recall that the problems of the formalism originate
from the mid-point insertions in the definitions of the gauge
transformations, not from the mid-point insertion in the definition of
the action. In fact, when string fields with mid-point insertions are
assumed to be outside the space of allowed string fields, the use of
mid-point insertions in the definition of the action does not lead to any
problems. We refer again to~\cite{Kroyter:2009zj} for further discussion on
this issue.

The modified cubic theory solved the problems of the original cubic theory
by working with NS fields in the ``neutral'' (zero) picture. This cannot be
imposed on the Ramond fields, since they carry half-integer
picture numbers. The closest one can get is to use a $\pm\frac{1}{2}$
picture. Of these two options, the ``natural'' Ramond ($-\frac{1}{2}$)
picture was selected. The fact that a non-zero string field is used
implies that picture changing operators appear in the equations of
motion and gauge transformations.
These operators must be inserted at the mid-point, leading to
potential singularities of the theory. If two Ramond fields are allowed,
one can write an action, whose equations of motion will not include
picture changing operators other than the global $Y_{-2}$ insertion.
Of course, the physical theory has only one Ramond field. Hence, a
constraint should be imposed. This constraint is bound to include picture
changing operators, leading to the usual equations of motion~(\ref{CubEOM}).
So one may be under the impression that nothing is gained. Still, the
fact that the new action is equivalent (as we shall promptly see) to
the unconstrained non-polynomial action may suggest that these two actions
might have some meaning even before the constraint is imposed. Moreover,
one might speculate that the new action can be useful for quantization
and for generalizations of the theory.

From the above discussion one can immediately guess the action\footnote{The
proofs of the various properties of the mapping between this action and the
unconstrained non-polynomial action are quite similar to those
of~\ref{sec:main}. Hence, we tend to be brief here.},
\begin{equation}
\label{2RCubaction}
S=-\int Y_{-2}\Big(\frac{1}{2}A Q A+\frac{1}{3}A^3 +
   \frac{1}{2} \tilde \al Q \al + \frac{1}{2}A [\al,\tilde\al]\Big)\,,
\end{equation}
with $\al$ having picture number $n_p(\al)=-\frac{1}{2}$ as before and
$n_p(\tilde \al)=\frac{1}{2}$.
From this action follow the equations of motion (omitting the global
$Y_{-2}$),
\begin{subequations}
\label{EOM2R}
\begin{align}
\label{2RAEOM}
QA+A^2+\frac{1}{2} [\al,\tilde\al] &= 0\,,\\
\label{2RalEOM}
Q\al+ [A,\al] &= 0\,,\\
\label{TilAlEOM}
Q\tilde\al+[A,\tilde\al] &= 0\,.
\end{align}
\end{subequations}

It is interesting to notice that the three string fields can be
unified in terms of a single string field simply by adding them.
Define,
\begin{equation}
\hat A=A+\frac{\al+\tilde\al}{\sqrt{2}}\,,
\end{equation}
and expand in terms of these constituents the natural action for $\hat A$,
\begin{equation}
\label{compactAction}
S=-\int Y_{-2}\Big(\frac{1}{2}\hat A Q\hat A+\frac{1}{3}\hat A^3\Big).
\end{equation}
Expanding the integrand one gets the integrand of the action~(\ref{2RAEOM})
together with some other terms. However, all other terms have wrong
picture numbers and hence can be safely dropped out of the action.

The constraint we should impose is,
\begin{equation}
\label{cubConst}
\tilde \al=X\al \quad \Longleftrightarrow \quad \al=Y\tilde \al\,.
\end{equation}
While both representations above are correct, the left one is more
accurate in the sense that $\tilde \al$ is the string field with the
actual mid-point $X$ insertion. A genuine $Y$ insertion is
prohibited, since it would lead to singularities with the
$Y_{-2}$ insertion in the action.
Keeping this constraint in mind, we also prohibit explicit
$X$ insertions in $\al$.
Applying this constraint, the action and the equations of motion reduce
to~(\ref{cubAction}) and~(\ref{CubEOM}) respectively.

At the linearized level the action~(\ref{2RCubaction}) in invariant under
three independent gauge transformations,
\begin{equation}
\delta A=Q\La\,,\qquad \delta \al=Q\chi\,,\qquad
    \delta \tilde \al=Q\tilde \chi\,. 
\end{equation}
However, only the first of these gauge transformations has an extension at
the non-linearized level. Hence, the complete gauge invariance of the theory
reads,
\begin{subequations}
\label{CubGauge2Fer}
\begin{align}
\delta A=&\, Q\La+[A,\La]\,,\\
\delta \al=&\, [\al,\La]\,,\\
\delta \tilde\al=&\, [\tilde\al,\La]\,.
\end{align}
\end{subequations}
This can be understood from the compact form of the
action~(\ref{compactAction}). This form implies that the gauge symmetry is,
\begin{equation}
\hat A \rightarrow e^{-\hat\La}(Q+\hat A)e^{\hat \La}\,.
\end{equation}
If one tries to substitute into $\hat \La$ any component, whose picture
is non-zero, it will result in taking the string field $\hat A$ out of the
allowed picture-number range, $-\frac{1}{2}\leq p\leq \frac{1}{2}$.
Hence, $\hat \La$ should be restricted to have only the zero picture
component $\La$ and the gauge transformation reduces (in its linearized form)
to~(\ref{CubGauge2Fer}).

Imposing the constraint~(\ref{cubConst}) leads to an enhancement of the gauge
symmetry to~(\ref{cubGauge}) (modulo the consistency
problem of this gauge transformation), in analogy with the situation in the
non-polynomial theory.
In fact, the absence of gauge symmetries for the fermionic string fields
should be expected, if we indeed believe (and shortly prove) that this
theory is equivalent to the non-constrained non-polynomial theory.
This can be seen by counting degrees of freedom.
The non-polynomial theory resides in the large Hilbert space, that has
double the degrees of freedom of the small Hilbert space, due to the
presence of the $\xi_0$ mode. In this space both $Q$ and $\eta_0$ are trivial
and hence gauge transformations based on them reduce the degrees
of freedom by a half. For the boson field, this theory has the $\La_1$ gauge
symmetry that effectively implies that the degrees of freedom of the theory
are isomorphic to those of the small Hilbert space. Then, on top of this
gauge symmetry there is also the $\La_0$ gauge symmetry that reduces the
degrees of freedom of the theory in exactly the same way that $\La$
reduces those of the cubic theory. Note that we can no longer claim that
this is a reduction by ``a half'', since in the small Hilbert space $Q$ is
no longer trivial. The two fermionic gauge symmetries of the theory, namely
$\La_{-\frac{1}{2}}$ and $\La_{\frac{3}{2}}$ reduce to a half the degrees
of freedom of $\Xi$ and $\Psi$ respectively, rendering them potentially
equivalent to $\al$ and $\tilde \al$. Had the cubic theory had more gauge
symmetry, its degrees of freedom could not have matched those of the
non-polynomial one.

Our goal now is to find the mapping between the two theories.
We propose the mapping,
\begin{equation}
\label{2RMap}
\Phi=PA\,,\qquad \Psi=-i P\tilde \al\,,\qquad	\Xi=-iP \al\,.
\end{equation}
We have to show that under this mapping solutions of the equations
of motion~(\ref{EOM2R}) are mapped to solutions of~(\ref{NPeqNoConstr}).
From~(\ref{2RMap}) we find that the l.h.s
of~(\ref{NPeqPhi}) is given by,
\begin{equation}
\eta_0 (e^{-\Phi} Q e^\Phi)+
  \frac{1}{2}[\eta_0 \Psi, e^{-\Phi} Q \Xi e^\Phi]=-Y\big(QA+A^2
    +\frac{1}{2}[\al,\tilde \al]\big),
\end{equation}
which vanishes in light of the equation of motion~(\ref{2RAEOM}).
Then, we find,
\begin{equation}
e^{-\Phi}Q\Xi e^\Phi=-i\al\,,
\end{equation}
where we used~(\ref{PP}). This implies that~(\ref{NPeqXi})
holds.

For calculating~(\ref{NPeqPsi}), define
\begin{equation}
\hat \al\equiv -i Y \tilde \al\,,
\end{equation}
and evaluate,
\begin{equation}
\label{FirstCol}
Q\big(e^\Phi\eta_0\Psi e^{-\Phi}\big)=Q\big((1+PA)\hat \al (1-PA)\big)=
Q\hat \al + [A,\hat\al] - P Q[A,\hat\al]=0\,,
\end{equation}
where the last equality follows from~(\ref{TilAlEOM}).
The last manipulation can be criticized on the ground that a factor
of $Y$ was ``hidden'' in the definition of $\hat \al$. When considered
explicitly this factor would produce divergences with the factors of $P$
that multiply it. We can avoid this problem in one of two ways.
One way is to recall that when we enforce the constraint
relating $\al$ and $\tilde \al$ the later has an explicit factor of
$X$ multiplying it. We can declare that $\tilde \al$ has
to have such an insertion even without the constraint. This does not change
a priori the amount of degrees of freedom of $\tilde \al$ and solves the
problem. Nonetheless, this resolution is not quite satisfactory since
our aim was to obtain a theory, which at least a priori is free from
explicit mid-point insertions over string fields.
The other way is to rely on the fact that we have to regularize the mappings
anyway, in order to produce a sensible action.
Then, we can declare that obtaining the above
result without any finite corrections is a symmetry principle for the
regularization scheme. This resolution further constrains
the needed regularization. However, we have more string fields now in our
disposal, so it is plausible that a regularization exists.

For the inverse mapping we choose,
\begin{subequations}
\label{invMap2R}
\begin{align}
\label{alOfXi}
\al &= ie^{-\Phi}Q\Xi e^\Phi\,,\\
\label{TilAlOfPsi}
\tilde \al &= i\eta_0 X \Psi\,,\\
A &= e^{-\Phi} Q e^\Phi + \frac{1}{2}\xi
 [\eta_0 \Psi, e^{-\Phi} Q \Xi e^\Phi]\,.
\end{align}
\end{subequations}
The equations of motion of the non-polynomial theory~(\ref{NPeqNoConstr})
imply that $A$, $\al$ and $\tilde \al$ live in the small Hilbert space as
they should.
Next, we have to verify that the equations of motion~(\ref{EOM2R})
also follow from~(\ref{NPeqNoConstr}) and the mapping~(\ref{invMap2R}).
The proof for~(\ref{2RAEOM}) and~(\ref{2RalEOM}) is straightforward.
For the evaluation of the last equation~(\ref{TilAlEOM}), define
\begin{equation}
\hat \Psi=i X\Psi\,.
\end{equation}
We then get,
\begin{align}
Q\tilde \al+[A,\tilde \al]&=\, Q\eta_0 \hat \Psi+\big[e^{-\Phi}Q e^\Phi+
  \frac{1}{2}\xi[\eta_0 \Psi,e^{-\Phi}Q \Xi e^\Phi],\eta_0 \hat\Psi\big]\\
\nonumber
 &=\,Q\eta_0 \hat \Psi+\big[(1-\xi \eta_0)e^{-\Phi}Q e^\Phi,\eta_0 \hat\Psi]=
  Q\eta_0 \hat \Psi-\eta_0 \xi Q\eta_0 \hat\Psi=\xi \eta_0 Q \eta_0 \hat \Psi
    =0\,,
\end{align}
where in the second equality~(\ref{NPeqPhi}) was used and in the next
equality~(\ref{NPeqPsi}) was used.
Note, that similarly to what we had in~(\ref{FirstCol}), a singularity
due to the collision of $X$ and $\xi$ is hidden in the definition of
$\hat \Psi$. Here, we have no excuse for claiming that $\Psi$ should always
have a factor $Y$ in its definition for cancelling the $X$ that
multiplies it. Hence, the only way out is to require the existence of a good
regularization scheme. We stress again, that as
in the previous cases where we relied on the existence of the regularization,
i.e., in the evaluation of the action in~\cite{Fuchs:2008zx} and
in~(\ref{FirstCol}), we have neither an explicit form of the regularization
nor a proof of its existence. We return to this point in
section~\ref{sec:conc}.

It is straightforward to see that composing the mapping~(\ref{invMap2R})
on~(\ref{2RMap}) results in the identity mapping of the cubic theory.
Suppose now that we compose the mappings in the opposite order. We expect to
get a finite gauge transformation. In fact we get,
\begin{subequations}
\label{2FerGaugeTrans}
\begin{equation}
e^\Phi_{\op{new}}=e^{-QP\Phi} e^\Phi\,,
\end{equation}
as before, while for the fermionic fields we get,
\begin{align}
\Psi_{\op{new}}=&\, \Psi-\eta_0 \xi \Psi\,,\\
\Xi_{\op{new}}=&\, P e^{-\Phi}Q \Xi e^\Phi=PQPe^{-\Phi}Q \Xi QP e^\Phi=
  e^{-QP \Phi}(\Xi-QP\Xi) e^{QP\Phi} \,.
\end{align}
\end{subequations}
Exponentiating the various transformations of~(\ref{actionGauge}) in order
to get the form of the finite gauge transformations one can see
that~(\ref{2FerGaugeTrans}) can be obtained by performing the following
finite gauge transformations in the order they are written,
\begin{equation}
\La_{\frac{3}{2}}=-\xi\Psi\,, \qquad \La_{-\frac{1}{2}}=-P\Xi\,, \qquad
\La_0=-P\Phi\,.
\end{equation}

Let there be two gauge equivalent solutions of the non-polynomial theory.
The infinitesimal gauge transformations are generated by the four gauge
fields $\La_p$ with $p=-\frac{1}{2},0,1,\frac{3}{2}$. Considering each of
these transformations and mapping it to the cubic theory we see that only
$\La_1$ induces a non-trivial transformation of the cubic string fields.
This variation is given by~(\ref{CubGauge2Fer}), with the identification
\begin{equation}
\La=\eta_0\La_1\,.
\end{equation}
We conclude that gauge equivalent configurations are mapped to gauge
equivalent configurations in this direction.
In the side of the cubic theory we have only one gauge transformation to
consider. Substituting~(\ref{CubGauge2Fer}) into the mapping gives,
\begin{equation}
\delta e^\Phi=e^\Phi\La-QP\La e^\Phi\,,\qquad \delta\Psi=[\Psi,\La]\,,
   \qquad \delta\Xi=[\Xi,QP\La]\,.
\end{equation}
This is a gauge transformation in the side of the non-polynomial theory
with the gauge string fields given by,
\begin{equation}
\La_0=-P\La\,,\qquad \La_1=\xi \La\,,\qquad
\La_{-\frac{1}{2}}=\La_{\frac{3}{2}}=0\,.
\end{equation}
We can now conclude that, when no constraints are imposed, the mapping of
gauge orbits between the two-Ramond-field
cubic theory and the unconstrained non-polynomial theory is bijective.

The proof of the equality of the action of solutions in both theories is
very similar to what we presented in~\ref{sec:action}. Again, we prove that
the Ramond-sector contribution to the action is zero for solutions in both
theories. For the non-polynomial theory the proof only used~(\ref{NPeqXi}),
which does not depend on the constraint. Hence, there is nothing new to prove
here. For the cubic theory the Ramond part
of the action integrand is bi-linear in $\al$ and $\tilde \al$, which
implies that it is equal to the star product of $\al$ with its equation of
motion, as well as to the star product of $\tilde \al$ with its equation
of motion. Any one of this equalities is enough to conclude that the action
of solutions gets no contribution from this sector.

We proved the equivalence of gauge orbits, which also implies the identity
of the cohomologies around solutions as well as the equality of the
action of corresponding solutions.
Hence, we conclude that the two theories are classically equivalent.

Finally, let us illustrate that the constraints one has to impose on both
theories are equivalent. Starting at the cubic theory we have to
impose~(\ref{cubConst}). This immediately implies,
\begin{equation}
\eta_0 \Psi=-iY\tilde\al=-i\al=e^{-\Phi}Q\Xi e^\Phi\,,
\end{equation}
which is just the constraint of the non-polynomial theory~(\ref{constraint}).
The other direction is as straightforward.
The equivalence of the constraints together with the other results of this
section give an ``alternative'' derivation of all the results of
section~\ref{sec:main}, since imposing the constraints on both theories
reduce them to the theories studied there.

%====================
\section{Conclusions}
%====================
\label{sec:conc}

In this work we proved the inconsistency of the modified cubic superstring
field theory. The inconsistency stems from collisions of picture changing
operators in the finite
form of its Ramond-sector gauge transformations. This state of affairs
implies that the cubic theory should either be abandoned or modified.
We believe that the later is the more sensible option, for two reasons. The
first, stressed throughout this paper, is the formal equivalence between this
theory and the non-polynomial theory. The second reason is the success of
this theory in describing the NS sector. In particular vacuum solutions
and dynamical tachyon condensation were studied both numerically and
analytically, with very impressive
results~\cite{Aref'eva:2000mb,Ohmori:2003vq,Erler:2007xt,Aref'eva:2008ad,
Fuchs:2008zx,Calcagni:2009tx}.
This can be compared to Witten's superstring field theory that failed to
reproduce any such results~\cite{DeSmet:2000je}.
We believe that had the cubic theory been completely wrong, it would have
not produced these results even in the NS sector. Hence, we need a
theory whose NS part reduces to the NS sector of the cubic theory in some
limit. This theory should also consistently include the Ramond sector.
A possible modification for the cubic theory was
recently proposed using non-minimal
sectors~\cite{Berkovits:2009gi,Kroyter:2009zj}. It seems, however, that
it cannot solve the consistency problem of the Ramond
sector~\cite{Kroyter:2009zj}.
In section~\ref{sec:2FermiCub}, we introduced a first step towards a
different sort of a modification. There, the Ramond sector was doubled in
order to avoid mid-point insertions of picture changing operators on
string fields.
The doubling of the Ramond sector kills supersymmetry. In order to restore
it, one might consider doubling the NS sector as well.
One might even speculate that such a doubling may be useful for
constructing closed superstring field theories, especially in light
of~\cite{Hull:2009mi}.

At any rate, the cubic theory with doubled Ramond sector is by itself not
satisfactory, since it does not have the correct amount of degrees of
freedom.
Following the example of~\cite{Michishita:2004by},
we tried to resolve this problem by introducing a constraint. It seems,
however, that such a constraint is bound to include
explicit mid-point insertions, which we have to avoid. Hence, we
have to look for another sort of resolution. An appealing possibility is
to further enlarge the field content as well as the gauge symmetry, such
that the final amount of degrees of freedom is reduced to the correct
one. We currently study this possibility and generalizations thereof.

The second issue studied in this work is the formal equivalence of the cubic
and non-polynomial theories. All the properties studied in this work were
shown to be invariant under our mappings, supporting the equivalence.
However, there is one more invariant that one might wish to consider. This is
the boundary state constructed from the solution~\cite{Kiermaier:2008qu}
(see also~\cite{Hashimoto:2001sm,Gaiotto:2001ji,Ellwood:2008jh,Kawano:2008ry,
Kishimoto:2008zj}).
To study this in the context of our equivalence one would first have to
combine some ideas from~\cite{Michishita:2004rx} and~\cite{Kiermaier:2008qu},
in order to define boundary states for the supersymmetric theories.
We currently study this subject.

It might seem strange that the Ramond parts of our mappings~(\ref{alOfPsi})
and~(\ref{PsiofAl}), include factors of $i$.
One might worry that our construction is inconsistent with
the reality condition of the string field. In fact, it is the other way
around. Recall that, in the Ramond sector, the coefficient fields are
Grassmann odd. The reality condition is obtained by composing Hermitian
conjugation and BPZ conjugation. While the first inverts the order of
insertions in the usual way, the second does not change the formal
Grassmann order~\cite{Taylor:2003gn}.
Hence, the reality of the string field $\al$ implies that
the string field $\xi \al$ is imaginary. The $i$'s take care just
of that.

Much of the recent renewal of interest in string field theory is due to
Schnabl's solution~\cite{Schnabl:2005gv} and subsequent
work~\cite{Okawa:2006vm,Fuchs:2006hw,Fuchs:2006an,Rastelli:2006ap,
Ellwood:2006ba,Fuchs:2006gs,Okawa:2006sn,
Erler:2006hw,Erler:2006ww,Schnabl:2007az,Kiermaier:2007ba,Erler:2007rh,
Okawa:2007ri,Fuchs:2007yy,Ellwood:2007xr,
Kishimoto:2007bb,Fuchs:2007gw,Kiermaier:2007vu,
Kiermaier:2007ki}.
Having explicit analytical expressions for string field theory solutions
evoked the realisation that these solutions can be formally represented as
pure gauge solutions.
In the bosonic theory, solutions can be written using singular gauge string
fields~\cite{Ellwood:2009zf}.
For the cubic superstring field theory in the NS
sector, the equivalence of~\cite{Fuchs:2008zx} provides a formal gauge form
for the solutions.
The ``gauge string field'' is formal in this case since it resides in the
large Hilbert space.
The extension of the equivalence to the Ramond sector
presented here does not seem to define a solution with a form of a formal
gauge solution.
This statement, however, is ill-defined, since the finite gauge
transformation of the Ramond sector does not exist for the cubic theory.
One might suspect that in a well defined refinement
of the cubic formalism it would be possible to write solutions as formal
gauge ones. However, in the cubic formalism with two Ramond
fields that we introduced in~\ref{sec:2FermiCub}, one sees that the gauge
transformation of the Ramond fields~(\ref{invMap2R}) is zero for a solution
with zero Ramond fields. Thus, one cannot get a solution with non-trivial
Ramond fields as a gauge solution around the vacuum. This can be traced to
the fact that we have no gauge symmetry in this theory whose generators
are fermionic. There is also no supersymmetry in this theory. It might be
the case that in a more physical refinement of the cubic theory there
will be a natural way for writing solutions as formal gauge ones.

While studying the mappings between the cubic and non-polynomial
two-Ramond-field theories, we got twice expressions, which were formally of
the form of a zero times a divergence that came from a mid-point collision
of operators. This implies that a regularization is needed in which these
expressions could be consistently set to zero. We thus have two requirements
from a consistent regularization, on top of the requirement that we had
from calculating the NS action~\cite{Fuchs:2008zx}. On the other hand, we
also have two more string fields, i.e., the Ramond ones and two more
mappings (in each direction) to
modify for defining the regularizations. Hence, the existence of a
sound regularization is as plausible as it is for the NS case. We would like
to stress that even regardless of the issue of singularities, a
regularization is desirable, since the mappings we introduced include
various mid-point insertions on string fields. Mid-point insertions on
string fields are highly constrained and a formulation that avoids them
altogether would be more reliable~\cite{Kroyter:2009zj}.

%=========================
\section*{Acknowledgments}
%=========================

I would like to thank Nathan Berkovits, Ted Erler, Udi Fuchs,
Michael Kiermaier, Leonardo Rastelli, Scott Yost and Barton Zwiebach for many
discussions on the issues covered in this work.

It is a pleasure to thank the organizers and participants of the KITP
workshop ``Fundamental Aspects of Superstring Theory'', where part
of this work was performed, for hospitality and for providing a very
stimulating and enjoyable environment. While at the KITP, this research was
supported by the National Science Foundation under Grant
No. PHY05-51164.
It is likewise a pleasure to thank the Simons Center for Geometry and Physics
and the organizers and participants of the ``Simons Workshop on String Field
Theory'' for a great hospitality and for many discussions on and around the
topics presented in this manuscript.

This work is supported by the U.S. Department of Energy
(D.O.E.) under cooperative research agreement DE-FG0205ER41360.
My research is supported by an Outgoing International Marie Curie
Fellowship of the European Community. The views presented in this work are
those of the author and do not necessarily reflect those of the European
Community.

\appendix
%==============================
\section{Quantum number tables}
%===============================
\label{app:picGh}

In this appendix we present the sector (R/NS), ghost number $n_g$, picture
number $n_p$ and parity of the string fields and string gauge fields.
Those of the modified cubic theory are
presented in table~\ref{tab:Cubic} while those of the non-polynomial theory
are presented in table~\ref{tab:NP}.
As the tables imply, string fields are odd and gauge string field are
even for the cubic theory, while the opposite is true for the non-polynomial
theory.
Also, in table~\ref{tab:Op} we present the quantum numbers of the operators
that compose the string fields.

\begin{table}[ht]
\begin{center}
\begin{tabular}{|c||c|r|r|c|}
\hline
string field & $R/NS$ & $n_g$ & $n_p$ & parity\\
\hline
$A$ & NS & 1 & 0 & 1\\
$\al$ & R & 1 & $-\frac{1}{2}$ & 1\\
$\La$ & NS & 0 & 0 & 0\\
$\chi$ & R & 0 & $-\frac{1}{2}$ & 0\\
\hline
\end{tabular}
\end{center}
\caption{Modified cubic theory: all string fields live in the small Hilbert
space.}
\label{tab:Cubic}
\end{table}
\begin{table}[ht]
\begin{center}
\begin{tabular}{|c||c|r|r|c|}
\hline
string field & $R/NS$ & $n_g$ & $n_p$ & parity\\
\hline
$\Phi$ & NS & 0 & 0 & 0\\
$\Psi$ & R & 0 & $\frac{1}{2}$ & 0\\
$\Xi$ & R & 0 & $-\frac{1}{2}$ & 0\\
$\La_{-\frac{1}{2}}$ & R & -1 & $-\frac{1}{2}$ & 1\\
$\La_0$ & NS & -1 & 0 & 1\\
$\La_{\frac{1}{2}}$ & R & -1 & $\frac{1}{2}$ & 1\\
$\La_1$ & NS & -1 & 1 & 1\\
$\La_{\frac{3}{2}}$ & R & -1 & $\frac{3}{2}$ & 1\\
\hline
\end{tabular}
\end{center}
\caption{Non-polynomial theory: string fields live in the large Hilbert
space. The subscripts of the gauge fields represent their picture
number.}
\label{tab:NP}
\end{table}
\begin{table}[ht]
\begin{center}
\begin{tabular}{|c||c|r|r|c|}
\hline
field & $h$ & $n_g$ & $n_p$ & parity\\
\hline
$\partial X^\mu$ & 1 & 0 & 0 & 0\\
$\psi^\mu$ & $\frac{1}{2}$ & 0 & 0 & 1\\
$b$ & 2 & -1 & 0 & 1\\
$c$ & -1 & 1 & 0 & 1\\
$\xi$ & 0 & -1 & 1 & 1\\
$\eta$ & 1 & 1 & -1 & 1\\
$e^{q\phi}$ & $-\frac{q(q+2)}{2}$ & 0 & $q$ & $q \mod 2$\\
$J_B$ & 1 & 1 & 0 & 1\\
$P$ & 0 & -1 & 0 & 1\\
$X$ & 0 & 0 & 1 & 0\\
$Y$ & 0 & 0 & -1 & 0\\
$X_2$ & 0 & 0 & 2 & 0\\
$Y_{-2}$ & 0 & 0 & -2 & 0\\
\hline
\end{tabular}
\end{center}
\caption{The conformal weight $h$, ghost number $n_g$, picture
number $n_p$ and parity of some conformal fields.}
\label{tab:Op}
\end{table}

%===================================================
\section{Why should the bosonic mapping be modified}
%===================================================
\label{App:Why}

The extra piece in the r.h.s of~(\ref{A}) may seem strange.
Here, we want to discuss its origin.

The equation of motion~(\ref{eomCubA}) is potentially problematic,
since it involves the mid-point operator insertion $X$.
It implies that either $A$ or $\al$ are allowed such an insertion.
If we assume that $A$ is not allowed to have mid-point insertions,
it follows that either $\al^2=0$, which is too restrictive, or
that $\al^2$ contains a factor of $Y$ that cancels the $X$.
For that to be the case it should be possible to write
\begin{equation}
\al=\cO^1 \al^{(1)}+\cO^2 \al^{(2)}\,,
\end{equation}
where $\cO^{1,2}$ are zero weight primaries inserted at the mid-point,
whose ghost numbers are opposite and whose picture numbers sum up to $-1$.
These operators should also obey the OPE's
\begin{equation}
\label{OPEforAlAl}
\cO^1 \cO^1 \sim 0\,,\qquad\cO^2 \cO^2 \sim 0\,,\qquad \cO^1 \cO^2 \sim Y\,.
\end{equation}
We failed to find such operators, both in the NS sector and in the Ramond
sector. In particular the choice $\cO^1=Y,\ \cO^2=1$
does not obey the above OPE's\footnote{One can imagine relaxing the first
two OPE's in~(\ref{OPEforAlAl}), so as to allow a larger algebra of
mid-point insertion operators. We did not manage to find a way to construct
that either.}.

Let us now assume that the mid-point insertion originates from $A$.
We can decompose $A$ as
\begin{equation}
\label{AxiDecompose}
A=A_0+\xi A_{-1}\,,
\end{equation}
where the subscript represents the picture-number. Substituting
into~(\ref{eomCubA}) and collecting the coefficients of $X$, $\xi$ and 1,
that should separately vanish, we get,
\begin{subequations}
\begin{align}
\label{A0eq}
QA_0+A_0^2=0\,,\\
\label{Am1ConEq}
A_{-1}+\al^2=0\,,\\
\label{Am1eq}
QA_{-1}+[A_0,A_{-1}]=0\,.
\end{align}
The equation for the Ramond field~(\ref{eomCubAl}) reduces now to,
\begin{equation}
\label{Aleq}
Q\al+[A_0,\al]=0\,.
\end{equation}
\end{subequations}
Equation~(\ref{A0eq}) states that $A_0$ is a flat connection,
while~(\ref{Am1eq}) and~(\ref{Aleq}) imply respectively that $A_{-1}$ and
$\al$ are ``covariantly constant'' with respect to $A_0$.
Equation~(\ref{Am1ConEq}) is a constraint equation relating $A_{-1}$ and
$\al$. With this constraint imposed, (\ref{Am1eq}) follows
from~(\ref{Aleq}). At this stage, one may think that $A_{-1}$ can be
discarded altogether, since it is fixed by $\al$. This is not quite the
case, since from its definition~(\ref{AxiDecompose}) we see that its
contribution to $A$ lives in the large Hilbert space. Since $A$ itself
lives in the small Hilbert space, it follows that this contribution
is needed in order to cancel other terms in $A_0$ that live in the large
Hilbert space.

Comparing~(\ref{AxiDecompose}) to~(\ref{A}), we see that
\begin{equation}
A_0=e^{-\Phi} Q e^\Phi\,.
\end{equation}
This is the expression for $A$ that one gets without the Ramond sector.
Adding the Ramond sector implies that this expression does not live anymore
in the small Hilbert space. The resolution is to define $A_{-1}$ as above,
in order to bring $A$ back to the small Hilbert space.
This amounts to using the mapping~(\ref{Map}).

\newpage
%=================
\bibliography{bib}
%=================

\end{document}